\documentclass[twocolumn,pra,aps,floatfix,longbibliography]{revtex4-1}
\usepackage{amsmath,esint}
\usepackage{amsfonts}
\usepackage{amssymb}
\usepackage{gensymb}
\usepackage{siunitx}
\usepackage{verbatim}
\usepackage{graphicx}
\usepackage{xfrac}
\usepackage{times}
\usepackage{natbib}
\usepackage[colorlinks=true,linkcolor=blue,urlcolor=blue,citecolor=blue]{hyperref}
\usepackage{dsfont}
\usepackage{xcolor}
\usepackage{url}
\usepackage{siunitx}
\usepackage{appendix}
\usepackage{multirow}
\usepackage{tabularx}

\newcolumntype{Y}{>{\centering\arraybackslash}X}
\begin{document}

\title{Optical $N$-insulators: topological obstructions in the atomistic susceptibility tensor}
\author{Todd Van Mechelen}
\affiliation{Purdue University, School of Electrical and Computer Engineering, Brick Nanotechnology Center, 47907, West Lafayette, Indiana, USA}
\author{Robert-Jan Slager}
\affiliation{TCM Group, Cavendish Laboratory, University of Cambridge, J.J.~Thomson Avenue, Cambridge CB3 0HE, United Kingdom}
\affiliation{Department of Physics, Harvard University, Cambridge, Massachusetts 02138, USA}
\author{Sathwik Bharadwaj}
\affiliation{Purdue University, School of Electrical and Computer Engineering, Brick Nanotechnology Center, 47907, West Lafayette, Indiana, USA}
\author{Zubin Jacob}
\email{zjacob@purdue.edu}
\affiliation{Purdue University, School of Electrical and Computer Engineering, Brick Nanotechnology Center, 47907, West Lafayette, Indiana, USA}

\begin{abstract}

A powerful result of topological band theory is that nontrivial phases manifest obstructions to constructing localized Wannier functions. In Chern insulators, it is impossible to construct Wannier functions that respect translational symmetry in both directions. Similarly, Wannier functions that respect time-reversal symmetry cannot be formed in quantum spin Hall insulators. This molecular orbital interpretation of topology has been enlightening and was recently extended to topological crystalline insulators which include obstructions tied to space group symmetries. In this article, we introduce a new class of two-dimensional topological materials known as optical $N$-insulators that possess obstructions to constructing localized molecular polarizabilities. The optical $N$-invariant $N\in\mathbb{Z}$ is the winding number of the atomistic susceptibility tensor $\chi$ and counts the number of singularities in the electromagnetic linear response theory. We decipher these singularities by analyzing the optical band structure of the material -- the eigenvectors of the susceptibility tensor -- which constitutes the collection of optical Bloch functions. The localized basis of these eigenvectors are optical Wannier functions which represent the molecular polarizabilities at different lattice sites. We prove that in a nontrivial optical phase $N\neq 0$, such a localized polarization basis is impossible to construct. Utilizing the mathematical machinery of $K$-theory, these optical $N$-phases are refined further to account for the underlying crystalline symmetries of the material, generating a classification of the topological electromagnetic phase of matter.

\end{abstract}

\maketitle 

\section{Introduction}

Upon the discovery of time-reversal invariant topological insulators (TIs), the past decades have witnessed a revival in band theory \cite{HasanKane10_RMP, Qi11_RMP}. While such topological phases were initially based on unitary and anti-unitary symmetries \cite{Kitaev2009,Ryu_2010}, e.g. time-reversal ($\mathcal{T}$) and particle-hole ($\mathcal{C}$) symmetry, recent developments have illuminated the role of the ever-present crystalline symmetries \cite{Slager2012}. These handles provide a more direct route for realizing topological phases, as well as giving rise to novel types of band topologies 
\cite{Kruthoff2017, Chiu2016, PhysRevB.90.165114, Holler18_PRB,PhysRevLett.108.106403, WannierFloquet}. Apart from these recent pursuits in electronics, topological band theory has also transformed the bosonic domains \cite{Xie2012,Sylvain2019}. Topological properties of photons \cite{Wu2015,Vanmechelen2019,Kim2020}, phonons \cite{SusstrunkE4767,Li2020}, plasmons \cite{Jin2017,vanmechelen2020optical}, excitons \cite{Klembt2018,Wenjing2020}, Cooper-pairs \cite{Masatoshi2017,Trang2020} and magnons \cite{Diaz2019,Wang2021} are under active investigation. In fact, crystalline symmetries are arguably even more crucial for bosons because they lack the time-reversal invariant of their fermionic counterparts. It is therefore critical to classify the possible states of matter according to their topological crystalline phases.

These so called topological crystalline insulators (TCIs) can possess surface states protected by geometric symmetries \cite{Fu2011}, as opposed to unitary and anti-unitary symmetries. One such TCI is tin telluride (SnTe), which has a Dirac cone surface band protected by mirror symmetry \cite{Tanaka2012}. On the other hand, the very existence of a surface may break the protecting symmetry of a TCI phase, denying any possibility of boundary states. In this scenario, the bulk material bears all the interesting topological properties -- the most important feature being obstructions to constructing exponentially localized Wannier functions. Topological obstructions to Wannier functions were first discovered in Chern insulators \cite{Thouless_1984}, relating to translational symmetry. A similar realization was made in quantum spin Hall (QSH) insulators that one cannot construct Wannier functions that respect time-reversal symmetry \cite{Soluyanov2011}. Soon thereafter, this molecular interpretation of topology was expanded to include TCIs which are connected to space group symmetries \cite{Po2017, Bradlyn2017}, bridging the classification principles of $K$-theory to real space \cite{Kruthoff2017}. That is, in nontrivial TCIs, the space group representations of a topological band structure are not realizable in a localized molecular basis. The discontinuity between these two pictures cannot be resolved without closing the band gap. 

Despite such fervent research in TCIs, few works have considered the polarizability of these peculiar molecular orbitals \cite{Resta1994,Essin2009}. Indeed, one question that has remained unanswered is: if the orbitals cannot be exponentially localized does that imply their polarizabilities cannot either? The current paradigm in topological condensed matter cannot answer this question as all electromagnetic fields are assumed static (DC). Polarizability requires an optical theory of matter that takes into account fluctuating electromagnetic waves (AC) and the corresponding electron transitions. In this article, we utilize state-of-the-art insights in band theory to formulate a new class of topological electromagnetic matter: optical $N$-insulators (ONIs). The optical $N$-invariant $N\in\mathbb{Z}$ is the winding number of the atomistic susceptibility tensor $\chi$ and counts the number singularities in the electromagnetic linear response theory. It was recently shown that graphene's viscous Hall fluid possesses these exotic singularities and hosts topologically protected edge plasmons \cite{vanmechelen2020optical}. However, the underlying honeycomb crystal structure of graphene was ignored since the analysis was limited to a long wavelength continuum theory. We make no such assumption here and develop a microscopic (ab initio) optical theory of matter starting from the many-body Schr\"{o}dinger equation. 

\begin{figure*}
    \centering
    \includegraphics[width=\linewidth]{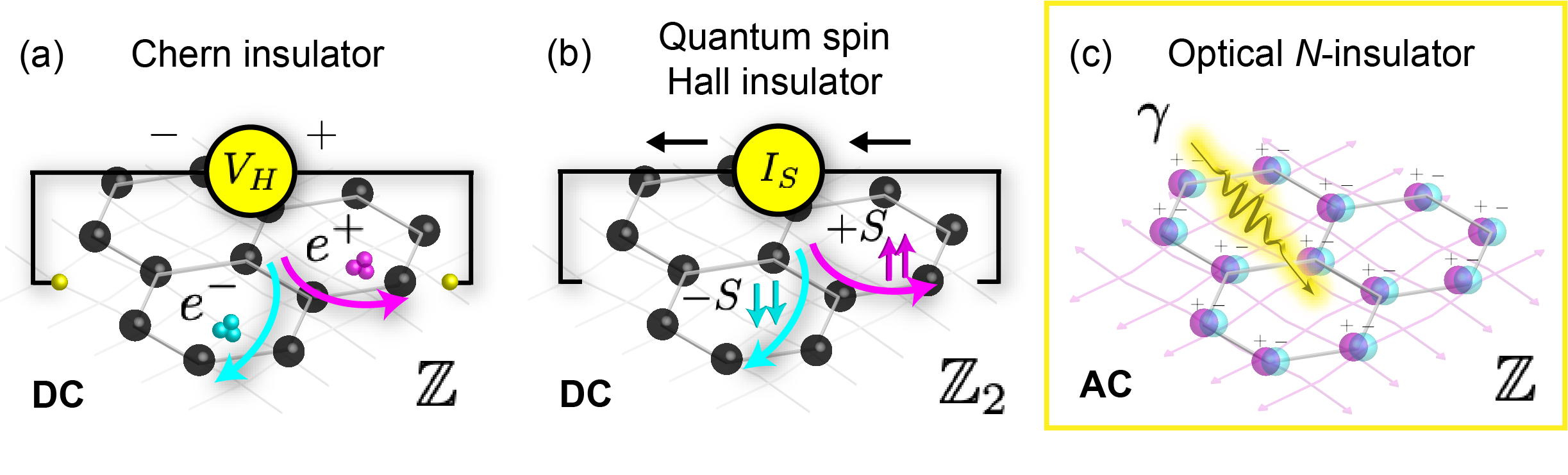}
    \caption{Topological phases in 2D. (a) The Chern phase $C\in\mathbb{Z}$ is connected to charge transport and generates a Hall voltage $V_H$ across the sample. $e^\pm$ are positive and negative charge carriers. (b) The quantum spin Hall (QSH) phase $\nu\in\mathbb{Z}$ is related to spin transport which creates a spin current $I_S$ across the sample. $\pm S$ are positive and negative spin carriers. (c) The optical $N$-phase $N\in\mathbb{Z}$ is associated with polarization transport. $\gamma$ is a photon and $\pm$ indicates the induced polarization of the electron cloud. Chern and QSH insulators are DC phases since they are defined for static electric fields. Optical $N$-insulators are AC phases as they must be interpreted for time-varying electric fields.}
    \label{fig:phases}
\end{figure*}

Utilizing the crystalline symmetry of the atomistic susceptibility tensor, we show that its eigenvectors are optical Bloch functions (OBFs) which constitute the optical band structure (OBS) of the crystal. We reveal that the $N$-invariant is interpreted as the geometric phase of the OBFs as they are parallel transported around the Brillouin zone. We then study the consequences of a nontrivial phase on molecular polarizabilities, which are represented by optical Wannier functions (OWFs). When $N\neq 0$, we prove it is impossible to construct exponentially localized OWFs that respect all symmetries of the system. We conclude the manuscript with a discussion on optical crystalline phases. Fortunately, the rigorous mathematical machinery of $K$-theory \cite{ATIYAH1966,Freed2013} can be exploited in electromagnetism which allows us to refine ONIs further to include the space group symmetries. Like TCIs, crystalline optical $N$-insulators (CONIs) can exhibit topological phases protected by geometric symmetries rather than unitary and anti-unitary symmetries. After analyzing the irreducible representations of the OBFs, we obtain a complete classification of the topological electromagnetic phase of matter. Figure~\ref{fig:phases} shows a comparison of 2D topological phases in condensed matter.

%ADD ONE PARA ON VISCOUS MCS THEORY AND QUANTUM GYRO-ELECTRIC EFFECT

%ADD ONE PARA ON GRAPHENE + HALL VISCOSITY, VANISHING MOKE, BANDGAP CLOSING with topological viscous plasmons, unique boundary conditions, LIMITATION THAT THIS PREVIOUS WORK WAS CONTINUUM THEORY, magneto-hydrodynamic Navier Stokes gave chi, new phases could exist by studying lattice symmetrices of chi-omega-q

\textbf{Note:} To avoid confusion, we adopt two vector conventions throughout the paper. We denote three-dimensional (3D) vectors with arrows $\Vec{\mathcal{A}}=(\mathcal{A}_x,\mathcal{A}_y,\mathcal{A}_z)$ and reserve boldface for two-dimensional (2D) vectors $\mathbf{A}=(A_x,A_y)$ which corresponds to strictly in-plane components.

\section{Atomistic Susceptibility Tensor}

Fundamentally, the transport of charge or polarization requires minimally coupling the system to an external gauge field $A_\mu$, which ensures local U(1) gauge symmetry. This is the foundational postulate of quantum electrodynamics. Assuming the spin response is negligible, the central object in linear response theory is the susceptibility tensor $\chi$, or equivalently the conductivity tensor $\sigma$. The susceptibility completely characterizes the transport properties of the material as it computes the induced polarization density $\vec{\mathcal{P}}$ due to an arbitrary electric field $\vec{\mathcal{E}}$,
\begin{equation}\label{eq:LinearResponseTheory}
\vec{\mathcal{P}}(\omega,\vec{r})=\int d\vec{r}' \chi(\omega;\vec{r},\vec{r}')\cdot\vec{\mathcal{E}}(\omega,\vec{r}').
\end{equation}
Note that $\vec{\mathcal{P}}$ and $\vec{\mathcal{E}}$ are microscopic fields -- no spatial averaging has been implemented. The susceptibility tensor $\chi$ is precisely the Green's function of the polarization density, and in a microscopic (ab initio) theory, $\chi$ has the symmetry of the atomic lattice. The general properties of $\chi$ are outlined in App.~\ref{app:Reponse}. At the atomistic level, the susceptibility is a highly dispersive and nonlocal object as it depends on the frequency $\omega$ and is generally nonzero even when $\vec{r}\neq\vec{r}'$. The photon energy $\omega$ lies within the electronic band gap of the insulator $0<\hbar\omega<E_\mathrm{bg}$, so the photon cannot be absorbed in the bulk -- it can only polarize the material. However, absorption may be permitted on the edge. Among other properties, the optical invariants of the bulk susceptibility predict whether gapless high-frequency (AC) edge currents can be stimulated by photons \cite{vanmechelen2020optical}.

Our focus is strongly correlated 2D materials but much of the formalism can be extended to 3D crystals, particularly the discussion on crystalline phases [Sec.~\ref{sec:SpaceGroups}]. From the Schr\"{o}dinger equation, we derive the zero temperature $T=0$ susceptibility tensor $\chi$, which represents the dielectric response of the many-body ground state $\psi_0$,
\begin{equation}\label{eq:AtomisticResponse}
\chi(\omega;\vec{r},\vec{r}')=-\frac{1}{\omega^2}\left[\zeta_0(\vec{r})\delta(\vec{r}-\vec{r}')\mathds{1}+\Sigma(\omega;\vec{r},\vec{r}')\right].
\end{equation}
$\mathds{1}$ is the $3\times 3$ identity matrix in Cartesian coordinates and $\delta$ is Dirac delta function. When $\omega$ is within the band gap, the response function is Hermitian and continuous, with continuous inverse. In this case, $\zeta_0$ is the instantaneous diamagnetic response and $\Sigma$ is the current-current correlation function that captures the paramagnetic response. These are defined unambiguously in App.~\ref{app:QuantumReponse}. It is imperative to remember that the linear response theory is only gauge invariant with $\zeta_0$ and $\Sigma$ taken together. 

We deduce that $\chi$ is periodic in the $x$-$y$ plane since the wave functions satisfy the Bloch condition $\hat{t}_\mathbf{R}|\psi_{n\mathbf{k}}\rangle=e^{i\mathbf{k}\cdot\mathbf{R}}|\psi_{n\mathbf{k}}\rangle$, where $\mathbf{k}$ is the crystal momentum of the many-body state [Eq.~(\ref{eq:BlochStates})]. The translation operator $\hat{t}_\mathbf{R}$ shifts all particles in the 2D crystal by an atomic lattice vector $\mathbf{R}$ which leaves the wave function unchanged up to a phase $e^{i\mathbf{k}\cdot\mathbf{R}}$. The susceptibility tensor is left invariant under a translation of the excitation point $\vec{r}$ and the source point $\vec{r}'$ by a lattice vector $\mathbf{R}$,
\begin{equation}\label{eq:PeriodicResponse}
\chi(\omega;\vec{r},\vec{r}')=\chi(\omega;\vec{r}+\mathbf{R},\vec{r}'+\mathbf{R}). 
\end{equation}
As a result, the in-plane photon momentum $\mathbf{q}$ is conserved up to a reciprocal lattice vector $\mathbf{G}\cdot\mathbf{R}\in 2\pi\mathbb{Z}$. We can therefore Fourier transform to the momentum space,
\begin{equation}\label{eq:TransformSusceptibility}
\chi(\omega;\vec{r},\vec{r}')=\sum_{\mathbf{q}\mathbf{q}'}\iint \frac{dq_z dq_z'}{(2\pi)^2}\chi(\omega;\vec{q},\vec{q}')e^{i\vec{q}\cdot\vec{r}}e^{-i\vec{q}'\cdot\vec{r}'}.
\end{equation}
Equation~(\ref{eq:PeriodicResponse}) implies $\chi(\omega;\vec{q},\vec{q}')$ is identically zero unless the in-plane momentum is conserved $\mathbf{q}-\mathbf{q}'=\mathbf{G}$, but the out-of-plane momenta $q_z-q_z'$ is arbitrary because there is no translational symmetry in this dimension. Hence, the susceptibility can be expressed as a matrix in the basis of $\mathbf{G}$ with $\mathbf{q}$ in the first Brillouin zone (FBZ) \cite{agranovich2013crystal}. Explicitly writing out all matrix indices with the excitation and source wave vectors defined as $\vec{q}=(\mathbf{q}+\mathbf{G},q_z)$ and $\vec{q}'=(\mathbf{q}+\mathbf{G}',q_z')$:
\begin{equation}
\begin{split}
\chi_{ij}(\omega;\vec{q},\vec{q}')&=\chi_{ij}(\omega;\mathbf{q}+\mathbf{G},q_z,\mathbf{q}+\mathbf{G}',q_z')\\&= \chi_{ij}^{\mathbf{G}\mathbf{G}'}(\omega;\mathbf{q},q_z,q_z').
\end{split}
\end{equation}
Subscripts denote tensor (polarization) degrees of freedom while superscripts denote lattice degrees of freedom $\mathbf{G}$. Our expression for the susceptibility in Eq.~(\ref{eq:TransformSusceptibility}) suppresses this cumbersome notation and will be adopted for the remainder of the paper. Converting Eq.~(\ref{eq:AtomisticResponse}) to the momentum space gives the compact expression for the susceptibility matrix,
\begin{equation}\label{eq:ResponseMomentum}
\chi(\omega;\vec{q},\vec{q}')=-\frac{1}{\omega^2}[\zeta_0(\vec{q},\vec{q}')\mathds{1}+\Sigma(\omega;\vec{q},\vec{q}')].
\end{equation}
The momentum space diamagnetic and paramagnetic responses are defined in App.~\ref{app:MomentumSpace}. The off-diagonal components $\mathbf{G}-\mathbf{G}'\neq \mathbf{0}$ are termed local-field effects which reflect the charge inhomogeneity of the crystal \cite{cohen_louie_2016}. A consequence is that long wavelength perturbations also give rise to short wavelength responses \cite{Wiser1963}. Local-field effects can be substantial such as the corrections to the Coulomb screening in semiconductors \cite{Zangwill1980} and optical rotatory power in $\alpha$-quartz \cite{Jonsson1996}. In this manuscript, we are concerned with the topological ramifications of these terms.

Components of the susceptibility matrix can be related to one another through complex conjugation. Regardless of the crystalline symmetry [Sec.~\ref{sec:SpaceGroups}], any susceptibility matrix has discrete anti-unitary symmetry since electromagnetism is a real-valued field theory. In the reciprocal space, $\chi$ satisfies the reality condition,
\begin{equation}\label{eq:Realfunction}
\chi(\omega;\vec{q},\vec{q}')=\chi^*(-\omega;-\vec{q},-\vec{q}').
\end{equation}
Equation~(\ref{eq:Realfunction}) implies electromagnetism belongs to universality class D -- the class of real bosons \cite{Ryu_2010}. In condensed matter, this is referred to as particle-hole symmetry ($\mathcal{C}$) and is normally associated with topological superconductors \cite{Onoeaaz8367,ZhongWang2012}. However, $\mathcal{C}$ symmetry is only approximate in superconductors but is an exact symmetry in electromagnetism. When considering variations in the eigenvalues $\omega$ and $\mathbf{q}$, the susceptibility matrix represents a mapping from the 2+1D momentum space to the general real linear group $\mathrm{GL}(D_f,\mathbb{R})$. $D_f$ denotes the degrees of freedom -- i.e. the total $D_f\times D_f$ matrix dimension of $\chi$, which includes both tensor (polarization) and $\mathbf{G}$ indices. $D_f$ may be infinite in principle but will usually be cut off at some suitably large value of $\mathbf{G}$.

\begin{table}
\caption{Three distinct types of optical materials. (a) Dielectrics are capacitive with strictly positive eigenvalues $\lambda>0$. (b) Metals are inductive with strictly negative eigenvalues $\lambda<0$. (c) Hybrid materials have both positive and negative eigenvalues $\lambda \gtrless 0$. Only hybrid materials support topological electromagnetic phases.}
\label{tab:optical_phases}
\begin{tabularx}{\linewidth}{X|YYY}
\hline\hline
 Material & (a) Dielectric & (b) Metal & (c) Hybrid \\\hline
 Eigenvalues &$\lambda>0$ & $\lambda <0 $& $\lambda\gtrless 0$\\
Electric field &$\leftarrow \vec{\mathcal{E}}$ & $\leftarrow \vec{\mathcal{E}}$ & $\leftarrow \vec{\mathcal{E}}$\\
Polarization &$\leftarrow \vec{\mathcal{P}}$ & $\rightarrow \vec{\mathcal{P}}$ & $\leftrightarrow \vec{\mathcal{P}}$\\
Classification & trivial & trivial & nontrivial\\
\hline\hline
\end{tabularx}
\end{table}

\section{Atomistic optical band structure}

\subsection{Optical Bloch functions}

The degree of freedom $D_f$ is precisely the number of eigenvectors needed to diagonalize the susceptibility matrix $\chi$. This collection of eigenvectors $\vec{e}_\lambda$ are optical Bloch functions (OBFs) which represent the optical band structure (OBS) of the crystal. The concept of an optical (or dielectric) band structure was proposed many years ago to understand microscopic electronic screening and its dependence on crystalline symmetries \cite{Hybersten1987,BALDERESCHI1979131,Car1981,Cudazzo2011,Huser2013,Latini2015}. This approach is tremendously powerful for $GW$ calculations \cite{Wilson2008,Wilson2009} and is also referred to as the projective dielectric eigendecomposition (PDEP) technique \cite{Pham2013,Govoni2015}. However, the scope has been limited to the longitudinal response function as the main focus has generally been Coulomb interactions. In the optical regime $\omega\neq 0$ electric fields can be both longitudinal and transverse, so we must consider the full susceptibility matrix which includes the longitudinal, transverse, and gyrotropic (Hall) responses [Eq.~(\ref{eq:ResponseMomentum})]. Our main contribution is showing that these systems can be optically nontrivial and give rise to new topological electromagnetic phases of matter.

OBFs are the normal modes of the crystal and satisfy an eigenvalue equation. Upon an optical perturbation of the form $\vec{\mathcal{E}}\propto \vec{e}_\lambda$, the response of the material is proportional $\vec{\mathcal{P}}\propto \vec{e}_\lambda$ up to a screening factor,
\begin{subequations}\label{eq:EigenPolarization}
\begin{equation}
\lambda_{\omega\mathbf{q}}\vec{e}_{\lambda}(\omega,\vec{q})=\sum_{\mathbf{G}'}\int  \frac{dq_z'}{2\pi}\chi(\omega;\vec{q},\vec{q}') \cdot \vec{e}_{\lambda}(\omega,\vec{q}').
\end{equation}
The screening factors $\lambda_{\omega\mathbf{q}}$ denote the eigenvalues of the susceptibility matrix at a particular photon energy $0<\hbar\omega<E_\mathrm{bg}$ and momenta $\mathbf{q}$. It is imperative not to confuse the OBF index $\lambda$ with the electronic band index $n$ as they refer to fundamentally different quantities. The wave functions $\psi_n$ represent the stable atomic configurations of the crystal. The OBFs $\vec{e}_\lambda$, on the other hand, characterize the collective dipole moments due to fluctuations in these configurations. Since the OBFs $\vec{e}_\lambda$ form a complete orthonormal set, we may perform a spectral decomposition of $\chi$ with respect to these modes,
\begin{equation}\label{eq:SpectralDecomp}
\chi(\omega;\vec{q},\vec{q}') =\sum_{\lambda}^{D_f} \lambda_{\omega\mathbf{q}}\vec{e}_{\lambda}(\omega,\vec{q})\otimes\vec{e}^*_{\lambda}(\omega,\vec{q}').
\end{equation}
\end{subequations}
$\otimes$ denotes the Cartesian outer product. Positive modes $\lambda>0$ are capacitive because the induced polarization points in the same direction as an applied field. Negative modes $\lambda<0$ are inductive since the polarization is reversed. Topologically, the capacitive $\lambda>0$ and inductive $\lambda<0$ modes define two distinct sectors of $\chi$ that cannot be continuously deformed into one another. We emphasize this point as it is critical to the topological classification of OBSs. The existence of inductive modes $\lambda<0$ is fundamentally necessary to achieve nontrivial topologies. Inductive modes in a passive dielectric medium would be exotic but can arise from exchange and correlation effects \cite{Car1981}. Moreover, such modes routinely emerge in time-reversal broken media with Hall responses which is our main platform for topological electromagnetic phases of matter. Table~\ref{tab:optical_phases} lists the three distinct types of optical materials.

\subsection{Optical Wannier functions}

Analogous to Wannier functions for atomic wave functions $\psi_n$, the OBFs $\vec{e}_\lambda$ can be expanded in a localized state basis, as opposed to the extended Bloch state basis. This localized basis is understood as the molecular polarizabilities \cite{Hunt1,Hunt2} at different lattice sites $\mathbf{R}$, which we refer to as optical Wannier functions (OWFs). Unlike traditional electronic Wannier functions, the OWFs $\vec{f}_{\lambda\mathbf{R}}$ are vector fields that characterize the molecular polarizability of a localized particle distribution to a fluctuating electric field $\vec{\mathcal{E}}$. Note that our construction of the susceptibility in terms of atomic polarizabilities goes beyond the classical Clausius-Mossotti relation \cite{Rysselberghe1932} since a crystal is neither homogeneous nor isotropic. In fact, one of the central results of this article is that there can be topological obstructions to such a construction. In the following we only consider 2D OWFs but the concept is easily extended to 3D.

If there exists a complete set of OWFs $\vec{f}_{\lambda\mathbf{R}}$, that form a representation of the space group, then OBFs $\vec{e}_\lambda$ can always be constructed from these localized molecular polarizabilities (up to a gauge transformation),
\begin{equation}\label{eq:OWFs}
\vec{e}_\lambda(\omega,\vec{q})=\int_\mathrm{cell}d\vec{r} \sum_\mathbf{R}\vec{f}_{\lambda\mathbf{R}}(\omega,\vec{r})e^{-i\vec{q}\cdot(\vec{r}-\mathbf{R})}.
\end{equation}
The $\mathrm{cell}$ subscript implies integration over the primitive unit cell of the crystal. Since the OWFs $\vec{f}_{\lambda\mathbf{R}}$ are exponentially localized, the OBFs $\vec{e}_\lambda$ are necessarily analytic and single-valued throughout the entire Brillouin zone \cite{Brouder2007}. In the following Sec.~\ref{sec:Ninvariant} we will understand when OBFs cannot be made analytic and when the construction of OWFs fails.

\section{Optical $N$-insulators}\label{sec:Ninvariant}

In a recent article \cite{vanmechelen2020optical}, we predicted a new topological phase of matter in graphene that is intimately related to polarization transport. This optical $N$-phase is fundamentally different from the conventional Chern (TKNN) \cite{Haldane1988,Thouless1982} and spin Hall phases \cite{Kane2005} because it is a manifestation of the response function $\chi$, not the wave function $\psi$. Moreover, nontrivial phases only arose in the many-body electron fluid regime and therefore are expected in strongly correlated systems. A critical component of such materials is nonlocal (or viscous) Hall conductivity \cite{Berdyugin2019}. In graphene, Hall viscosity causes the representation of the chiral electron fluid to change at large momenta \cite{VanMechelen2020} which creates a singularity in the response function. Here, we generalize this idea to include local-field effects and the corresponding lattice symmetries.

Since the susceptibility matrix $\chi$ is a Green's function, it naturally possesses a topological invariant that is conserved under continuous deformations \cite{volovik2009universe,nonherzero,Gurarie2011,Slager_review,slager2015impurity}. The homotopy of $\chi$ is encapsulated in the following 2+1D optical $N$-invariant,
\begin{equation}\label{eq:Volovik}
N=\frac{\epsilon^{\alpha\beta\gamma}}{24\pi^2} \iint d\Omega d\mathbf{q}~\mathrm{tr}\left[\chi\frac{\partial\chi^{-1}}{\partial q_\alpha}\chi\frac{\partial\chi^{-1}}{\partial q_\beta}\chi\frac{\partial\chi^{-1}}{\partial q_\gamma}\right],
\end{equation}
where $q_\alpha=(\Omega,\mathbf{q})$ denotes the total momentum coordinate and $\omega\to\Omega$ has been analytically continued into the complex frequency plane $\Omega$. Note that the matrix product and trace in Eq.~(\ref{eq:Volovik}) involves summation over the tensor and Fourier elements $\mathbf{G}$, as well as the convolution over the out-of-plane momenta $q_z$ which is formally defined in App.~\ref{app:MomentumSpace}. The momentum integral $d\mathbf{q}$ spans the first Brillouin zone $\mathbf{q}\in\mathrm{FBZ}$. The temporal integral $d\Omega$ however, is performed vertically over all imaginary (Matsubara) frequencies
$\Omega\in(\omega-i\infty,\omega+i\infty)$. As before, the photon energy $\omega=\Re(\Omega)$ lies within the electronic band gap $0<\hbar\omega< E_\mathrm{bg}$. Utilizing the $f$-sum rule, we have proven that the integral is nondegenerate and convergent (see App.~\ref{app:ProofEQ9}). We have also shown that $N$ is immune to perturbations in the optical response $\chi\to \chi+\delta\chi$ and therefore topologically quantized $N\in\mathbb{Z}$. Also, it should be noted that the optical $N$-invariant can be calculated using the susceptibility $\chi$ or conductivity $\sigma$ matrices as they are homotopy equivalent. A proof is provided in App.~\ref{app:CondHomotopy}.

The existence of the optical $N$-invariant reflects the fact that $\pi_3[\mathrm{GL}(D_f,\mathbb{R})]=\mathbb{Z}$, which states that the third homotopy group of $\mathrm{GL}(D_f,\mathbb{R})$ is isomorphic to $\mathbb{Z}$. Nevertheless, the expression in Eq.~(\ref{eq:Volovik}) has severe disadvantages both practically and conceptually. It involves inverting a very large matrix and integrating the dynamical dispersion of $\chi$ over all Matsubara frequencies, which is a formidable task numerically. Utilizing the spectral decomposition of the OBS [Eq.~(\ref{eq:SpectralDecomp})], we prove that the $N$-invariant is equivalently calculated in terms of the geometric phase of the OBFs $\vec{e}_\lambda$,
\begin{subequations}\label{eq:BerryCurvature}
\begin{equation}
N=\frac{1}{2\pi}\int d\mathbf{q} ~\mathfrak{F} ,
\end{equation}
where $\mathfrak{F}=\sum_{\lambda>0}\mathfrak{F}_\lambda$ is the capacitive Berry curvature,
\begin{equation}
\mathfrak{F}(\omega,\mathbf{q})=-i\epsilon^{ij}\sum_{\lambda>0}\sum_\mathbf{G}\int \frac{dq_z}{2\pi}\frac{\partial \vec{e}_\lambda^*(\omega,\vec{q})}{\partial q_i}\cdot\frac{\partial \vec{e}_\lambda(\omega,\vec{q})}{\partial q_j}.
\end{equation}
\end{subequations}
It bears repeating that we may calculate $N$ for any value of $0<\hbar\omega< E_\mathrm{bg}$ in the band gap. The proof is presented in App.~\ref{app:ProofEq10}. We now have a very intuitive picture; the $N$-invariant corresponds to the parallel transport of the internal dipole moments around the Brillouin zone. Notice that the summation runs over all capacitive modes $\lambda>0$ but the existence of inductive modes $\lambda<0$ is paramount. If the response is purely capacitive/inductive then the topology is necessarily trivial because the sum over all OBFs vanishes $\sum_\lambda^{D_f}\mathfrak{F}_\lambda=0$.

\subsection{Topological obstructions to molecular polarizabilities}

One of the pioneering results of topological band theory, first revealed in the Chern phase \cite{Thouless_1984}, is that a nontrivial TKNN invariant implies it is impossible to construct Wannier functions that respect all symmetries of the crystal. In contemporary condensed matter physics, this is often considered the defining feature of a topological band structure \cite{Po2017, Bradlyn2017}. Analogously, the optical $N$-invariant represents a topological obstruction to OWFs and the construction of localized molecular polarizabilities. To demonstrate this explicitly, we assume it is possible to construct localized OWFs that from a representation of the translation group. Substituting the OWF expansion in Eq.~(\ref{eq:OWFs}) into Eq.~(\ref{eq:BerryCurvature}) gives,
\begin{equation}
N=-i\sum_{\lambda>0}\sum_\mathbf{R}\epsilon^{ij}R_iR_j\int_\mathrm{cell}d\vec{r}\left|\vec{f}_{\lambda\mathbf{R}}(\omega,\vec{r})\right|^2=0.
\end{equation}
The $N$-invariant is identically zero $N=0$, corresponding to the trivial phase. We arrive at a profound conclusion. If the OWFs $\vec{f}_{\lambda\mathbf{R}}$ are constructed from OBFs $\vec{e}_\lambda$ with a nontrivial $N$-invariant $N\neq 0$, they cannot fall off faster than $|\mathbf{r}-\mathbf{R}|^{-2}$ and have long-range extent. This is due to the fact that $\vec{e}_\lambda$ cannot be made an analytic and single-valued function of $\mathbf{q}$ throughout the entire Brillouin zone. Indeed, the condition that $\vec{f}_{\lambda\mathbf{R}}$ is exponentially localized requires a vanishing $N$-invariant $N=0$. Hence, there is an obstruction to constructing molecular polarizabilities in the atomic limit when $N\neq 0$.

\begin{table*}
\caption{Examples of 2D CONIs for the 17 wallpaper groups. Wallpaper groups $\Hat{G}$ are denoted in Hermann-Mauguin notation \cite{hahn2005international} and designate the spatial symmetry of the crystal. The second row lists the number of integers $\{N_i^\mathbf{q}\}$ that need to be specified in order to completely characterize the representation ($\mathcal{D}$) of the capacitive $\lambda>0$ OBFs $\vec{e}_\lambda$ within the specific wallpaper group. The third row signifies whether an optical $N$-invariant is present and the fourth row is the total number of integers needed to completely classify the topological phase of the OBS.}
\label{tab:SummaryPhases}
\begin{tabularx}{\textwidth}{lYYYYYYYYYYYYYYYYY}
\hline\hline
$\hat{G}$     & $p1$           & $p2$           & $pm$           & $pg$         & $cm$           & $p2mm$         & $p2mg$         & $p2gg$         & $c2mm$         & $p4$           & $p4mm$         & $p4gm$         & $p3$           & $p3m1$         & $p31m$         & $p6$            & $p6mm$         \\\hline
$\mathcal{D}$ & $\mathbb{Z}$   & $\mathbb{Z}^5$ & $\mathbb{Z}^3$ & $\mathbb{Z}$ & $\mathbb{Z}^2$ & $\mathbb{Z}^9$ & $\mathbb{Z}^4$ & $\mathbb{Z}^3$ & $\mathbb{Z}^6$ & $\mathbb{Z}^8$ & $\mathbb{Z}^9$ & $\mathbb{Z}^6$ & $\mathbb{Z}^7$ & $\mathbb{Z}^5$ & $\mathbb{Z}^5$ & $\mathbb{Z}^9$  & $\mathbb{Z}^8$ \\
 $N$           & $\mathbb{Z}$   & $\mathbb{Z}$   & 0              & 0            & 0              & 0              & 0              & 0              & 0              & $\mathbb{Z}$   & 0              & 0              & $\mathbb{Z}$   & 0              & 0              & $\mathbb{Z}$    & 0             
  \\
 Total         & $\mathbb{Z}^2$ & $\mathbb{Z}^6$ & $\mathbb{Z}^3$ & $\mathbb{Z}$ & $\mathbb{Z}^2$ & $\mathbb{Z}^9$ & $\mathbb{Z}^4$ & $\mathbb{Z}^3$ & $\mathbb{Z}^6$ & $\mathbb{Z}^9$ & $\mathbb{Z}^9$ & $\mathbb{Z}^6$ & $\mathbb{Z}^8$ & $\mathbb{Z}^5$ & $\mathbb{Z}^5$ & $\mathbb{Z}^{10}$ & $\mathbb{Z}^8$ \\\hline\hline
\end{tabularx}
\end{table*}

\section{Crystalline optical $N$- insulators}\label{sec:SpaceGroups}

We now direct our attention to the ever-present crystalline symmetries that refine ONIs further. This application of $K$-theory \cite{Kruthoff2017} predicts additional topological invariants and imposes constraints on the $N$-invariant itself. Although we only consider 2D crystalline materials here, the $K$-theory recipe can also be extended to 3D crystals. The importance of $K$-theory is expounded in its ability to predict topological phases when standard unitary and anti-unitary classifications would predict none. For instance, in space groups with mirror symmetry, the $N$-invariant vanishes but the material may still be topological in the sense of lacking an atomic limit. That is, there may still be obstructions to OWFs. Determining when this is the case is delicate and currently a matter of contention. Section~\ref{sec:GeneralPerspective} discusses the current paradigm in great detail. In this section, we only outline the topological invariants of the OBS that arise from a rigorous $K$-theory classification.

For 2D materials we examine the general layer space groups that make up all crystals periodic in two dimensions. The layer group is the 3D extension of the wallpaper group, which includes reflections in the third spatial dimension $z$. There are a total of 80 distinct layer groups $\hat{G}$ and the 17 wallpaper groups constitute a subgroup of these, which we use as instructive examples [Table~\ref{tab:SummaryPhases}]. We define an element $g\in \hat{G}$ of a particular space group as $g=\{R,\mathbf{a}\}$, where $R$ is an orthogonal matrix that describes rotations $\det(R)=1$ or reflections $\det(R)=-1$ of the axial crystallographic point group. $\mathbf{a}$ denotes translations in the $x$-$y$ plane. If the material belongs to a particular space group, the susceptibility matrix is invariant under all such elements $g$,
\begin{subequations}
\begin{equation}
\chi(\omega;\vec{q},\vec{q}')=R^{-1} \chi(\omega;R\cdot\vec{q},R\cdot\vec{q}') R ~e^{i(\mathbf{G}-\mathbf{G}')\cdot\mathbf{a}}.
\end{equation}
Since $\chi$ transforms under the little co-group of $\mathbf{q}$, the OBFs $\vec{e}_\lambda$ are naturally classified according to the irreducible representations (irreps) of that group. The action of a space group element on an OBF $\vec{e}_{\lambda}$ is defined as,
\begin{equation}
R^{-1}\cdot\vec{e}_{\lambda }(\omega,R\cdot\vec{q})e^{i\mathbf{G}\cdot\mathbf{a}}=\sum_{\lambda'} \mathcal{D}_{\lambda\lambda'}^{\mathbf{q}}(g)\vec{e}_{\lambda'}(\omega,\vec{q}).
\end{equation}
\end{subequations}
The point group element $R$ transforms an OBF with momentum $\vec{q}$ to another field with momentum $R\cdot\vec{q}$ and the same eigenvalue $\lambda_{\omega\mathbf{q}}=\lambda_{\omega R\cdot\mathbf{q}}$, while $\mathbf{a}$ translates the mode. The matrix $\mathcal{D}_{\lambda\lambda'}^{\mathbf{q}}(g)$ is a representation of the space group element $g=\{R,\mathbf{a}\}$ at the in-plane momentum $\mathbf{q}$. Space groups where $\mathbf{a}=\mathbf{R}$ is a pure lattice translation are called symmorphic and we need only consider ordinary representations of the point groups in these scenarios. On the other hand, nonsymmorphic space groups ($pg$, $p2mg$, $p2gg$, and $p4gm$ for example) contain glide planes where $\mathbf{a}$ is a fraction of a lattice translation. In this case projective representations need to be considered since translations cannot be separated from point group operations. These require special attention as there is an additional phase factor. The detailed procedure for nonsymmorphic space groups is outlined in App. A of Ref.~\cite{Kruthoff2017}.

\subsection{High-symmetry points}

At generic points in the fundamental domain (aka. the irreducible Brillouin zone), $\mathcal{D}$ is usually just a phase factor -- a one-dimensional representation. However, at certain high-symmetry points (HSPs) or high-symmetry lines (HSLs), the space group may permit higher dimensional representations (two, three, etc.). At a HSP, the momentum is invariant under a rotation or reflection $R\cdot\vec{q}=\vec{q}$. This means $R\cdot\vec{q}$ and $\vec{q}$ differ at most by a reciprocal lattice vector $\mathbf{G}$. All OBFs $\vec{e}_{\lambda}$ connected through the matrix $\mathcal{D}$ are necessarily degenerate at $\mathbf{q}$, where the degeneracy equals the dimension of the irrep $\dim(\mathcal{D})$ \cite{tinkham_group_2012}. The different possible irreps depends on the little co-group $G^\mathbf{q}$ at the momentum $\mathbf{q}$, which is the subgroup of all symmetry operations that keep a particular HSP or HSL fixed. In the absence of time-reversal symmetry, there are no additional constraints on the representations that may cause extra Kramers degeneracies. The representations can be real or complex and depend only the space group. %Note though, reality implies $\mathcal{D}_{\lambda\lambda'}^{\omega\mathbf{q}}(g)=[\mathcal{D}_{\lambda\lambda'}^{-\omega-\mathbf{q}}(g)]^*$ is self-conjugate which relates the representations for positive or negative photon energy -- the choice of which is a matter of preference.

Besides the optical $N$-invariant, the remaining topological invariants correspond to a set of integers $\{N^\mathbf{q}_i\}\in\mathbb{Z}$ that are related to the representations of the space group. 
\begin{quote}
In the spectral gap $0<\hbar\omega<E_\mathrm{bg}$, each integer $N^\mathbf{q}_i$ labels the number of capacitive $\lambda>0$ OBFs $\vec{e}_{\lambda}$ transforming under the $i$th irrep at the HSP $\mathbf{q}$. \end{quote}
The total amount of integers will differ depending on the space group as the number of irreps and HSPs is a property of the group itself. Furthermore, not all integers are independent as the representations must satisfy compatibility (gluing) conditions as proposed in Ref. \cite{Kruthoff2017}. Since one can traverse the Brillouin zone numerous ways, connecting HSPs along different HSLs, the eigenvalues associated with these operations must be mutually compatible. Thus, the gluing conditions relate the representations at different HSPs. Knowing the number of OBFs $\vec{e}_\lambda$ transforming under the irreps at a limited number of HSPs is sufficient to determine all representations. We reiterate that the set of integers $\{N_i^\mathbf{q}\}$ can be calculated for any $0<\hbar\omega<E_\mathrm{bg}$ within the electronic band gap. Table~\ref{tab:SummaryPhases} lists the total set of invariants needed to completely specify the optical crystalline phase in all 17 wallpaper groups. 

%In the following Sec.~\ref{sec:HallFluid}, we utilize this formalism to analyze the topology of the viscous Hall fluid.

\subsection{An example: $p2mm$}\label{sec:p2mm}

\begin{figure}
    \centering
    \includegraphics[width=\linewidth]{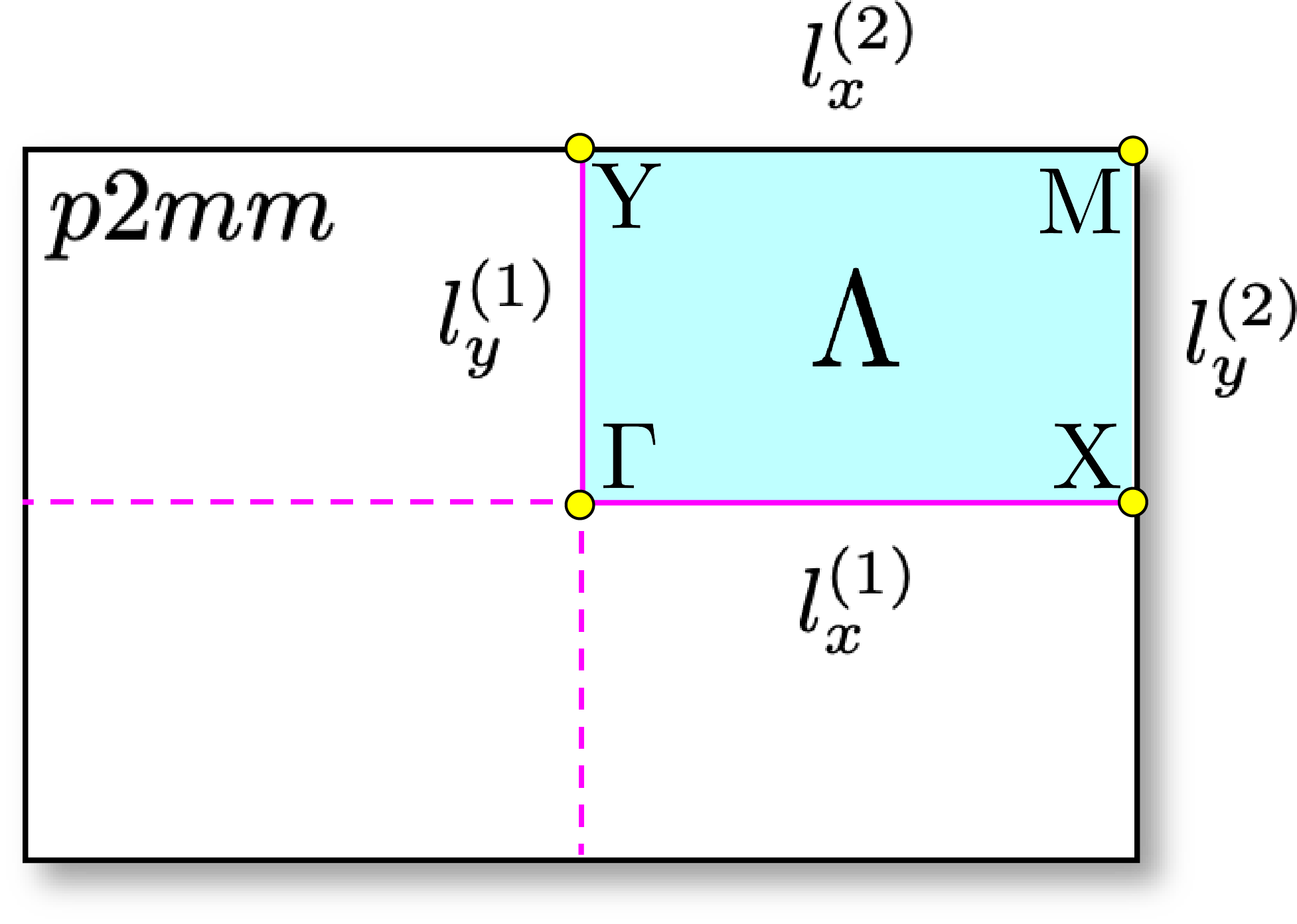}
    \caption{Brillouin zone of space group $p2mm$. The fundamental domain $\Lambda$ is indicated by the cyan region. All other momenta are related to $\Lambda$ by transformations in the dihedral point group $D_2$. This space group has four HSPs at $\Gamma$, $X$, $Y$ and $M$, connected by four HSLs $l_{x,y}^{(1,2)}$ which spans a rectangle.}
    \label{fig:p2mm}
\end{figure}

We address space group $p2mm$ as a simple but guiding example [Fig.~(\ref{fig:p2mm})]. A similar pedagogical demonstration is presented for $p4mm$ in Ref.~\cite{Kruthoff2017}. Space group $p2mm$ has two perpendicular planes of reflection symmetry corresponding to the dihedral point group $D_2=\mathbb{Z}_2\times\mathbb{Z}_2$. The operation of reflections $t_x$ and $t_y$ in either plane, takes the Bloch momentum $\mathbf{q}$ to,
\begin{equation}
t_x\cdot (q_x,q_y)=(-q_x,q_y), \qquad t_y\cdot (q_x,q_y)=(q_x,-q_y).
\end{equation}
Inversion is thus the combined operation $t_xt_y$ which gives all the operations of the space group. $p2mm$ is particularly straightforward as $D_2$ is an abelian group and therefore only has one-dimensional representations. $p2mm$ has four HSPs and a rectangular fundamental domain $\Lambda$ that is spanned by $\Gamma$, $X$, $Y$ and $M$. All HSPs feature the full $D_2$ symmetry which has four irreps [Table~\ref{tab:p2mm_groups}]. The character tables of the HSPs and HSLs are displayed in Tables~\ref{tab:GammaIrreps} and \ref{tab:HSLIrreps} respectively. Note, the optical $N$-invariant vanishes as the Berry curvature $\mathfrak{F}$ is odd under mirror symmetry. The OBS topology is governed entirely by the irreps of the capacitive modes at HSPs. To count the phases, we need to enumerate the number of unique OBF representations as not all are independent. 

We start with a set of $4\times 4=16$ integers $\{N_i^\mathbf{q}\}$ that label the number of irreps of the capacitive modes $\lambda>0$ at the four HSPs. The compatibility relations determine the constraints on $N_i^\mathbf{q}$ as the fields must retain their eigenvalues along the high-symmetry lines $l_{x,y}^{(1,2)}$. For instance, the number of even $N_+^{l_x^{(1)}}$ and odd $N_-^{l_x^{(1)}}$ modes along $l_{x}^{(1)}$ must equal the number of even and odd $t_y$ modes at $\Gamma$, 
\begin{subequations}
\begin{equation}
N_0^\Gamma+N_3^\Gamma=N_+^{l_x^{(1)}}, \qquad N_1^\Gamma+N_2^\Gamma=N_-^{l_x^{(1)}}.
\end{equation}
Similarly, $N_+^{l_x^{(1)}}$ and $N_-^{l_x^{(1)}}$ must equal the number of even and odd capacitive modes at $X$. Hence, we obtain a set of compatibility conditions that relates the even and odd modes at $\Gamma$ and $X$,
\begin{equation}
N_0^\Gamma+N_3^\Gamma=N_0^X+N_3^X, \qquad N_1^\Gamma+N_2^\Gamma=N_1^X+N_2^X.
\end{equation}
Following a similar procedure for the remaining HSPs, we obtain six more relations,
\begin{equation}
N_0^X+N_2^X=N_0^M+N_2^M, \qquad N_1^X+N_3^X=N_1^M+N_3^M,
\end{equation}
\begin{equation}
N_0^M+N_3^M=N_0^Y+N_3^Y, \qquad N_1^M+N_2^M=N_1^Y+N_2^Y,
\end{equation}
\begin{equation}
N_0^Y+N_2^Y=N_0^\Gamma+N_2^\Gamma, \qquad N_1^Y+N_3^Y=N_1^\Gamma+N_3^\Gamma.
\end{equation}
\end{subequations}
There are eight relations but only seven are independent, which denotes a rank of $7$. The total number of integers needed to completely specify the representations of the capacitive OBFs is therefore $16-7=9$. We conclude that the optical crystalline phase of $p2mm$, is classified by the elements of $\mathbb{Z}^9$. Again, due to mirror symmetry in space group $p2mm$, an $N$-invariant is absent $N=0$.

\begin{table}[htbp]
    \centering
        \caption{The little co-groups $G^\mathbf{q}$ within space group $p2mm$. These subgroups contain all symmetry operations that keep the momentum of a particular HSP or HSL fixed. The fundamental domain $\Lambda$ consists of the rectangular momentum space $0\leq q_{x,y} \leq \pi$ and its interior is denoted $\mathrm{int}(\Lambda)$.}
    \label{tab:p2mm_groups}
    \begin{tabularx}{\linewidth}{lYY}
        \hline\hline
    High-symmetry point     & $\mathbf{q}$ & Little co-group $G^\mathbf{q}$ \\\hline
    $\Gamma$   & $(0,0)$ & $D_2$\\
    $M$   & $(\pi,\pi)$ & $D_2$\\
    $X$   & $(\pi,0)$ & $D_2$ \\ 
    $Y$   & $(0,\pi)$ & $D_2$ \\
    $l_x^{(1)}$ & $(q_x,0)$& $\mathbb{Z}_2=\{1,t_y\}$\\
    $l_y^{(1)}$ & $(0,q_y)$& $\mathbb{Z}_2=\{1,t_x\}$\\
    $l_x^{(2)}$ & $(q_x,\pi)$& $\mathbb{Z}_2=\{1,t_y\}$\\
    $l_y^{(2)}$ & $(\pi,q_x)$& $\mathbb{Z}_2=\{1,t_x\}$\\
    $\mathrm{int}(\Lambda)$ & $(q_x,q_y)$& $\{1\}$\\
        \hline\hline
    \end{tabularx}
\end{table}

\begin{table}[htbp]
    \centering
        \caption{Character table of the dihedral point group $D_2$ for HSP $\Gamma$. The irreps are denoted by $\mathcal{D}_i$ and the columns are labeled by the conjugacy classes. All HSPs $\Gamma$, $X$, $Y$, and $M$ have little co-groups $D_2=\mathbb{Z}_2\times\mathbb{Z}_2$.}
    \label{tab:GammaIrreps}
    \begin{tabularx}{\linewidth}{lYYYY}
    \hline\hline
   $D_2$ & $\{1\}$ & $\{t_xt_y\}$ & $\{t_x\}$ & $\{t_y\}$\\\hline
   $\mathcal{D}_0$   & 1& 1& 1 &1  \\
   $\mathcal{D}_1$   & 1& 1& -1 &-1  \\
   $\mathcal{D}_2$   & 1& -1& 1 &-1  \\
   $\mathcal{D}_3$   & 1& -1& -1 &1  \\\hline\hline
    \end{tabularx}

\end{table}

\begin{table}[htbp]
    \centering
    \caption{Character table of $\mathbb{Z}_2$ for the HSLs $l_{x.y}^{(1,2)}$. Irreps are denoted by $\pm$ which corresponds to even or odd under reflection symmetry.}
    \label{tab:HSLIrreps}
    \begin{tabularx}{0.45\linewidth}{lYY}
    \hline\hline
   $\mathbb{Z}_2$ & $\{1\}$ & $\{t_y\}$ \\\hline
   $l_{x,+}^{(1,2)}$   & 1& 1  \\
   $l_{x,-}^{(1,2)}$   & 1 &-1 \\\hline\hline
    \end{tabularx}
       \qquad
        \begin{tabularx}{0.45\linewidth}{lYY}
    \hline\hline
   $\mathbb{Z}_2$ & $\{1\}$ & $\{t_x\}$ \\\hline
   $l_{y,+}^{(1,2)}$   & 1& 1  \\
   $l_{y,-}^{(1,2)}$   & 1 &-1 \\\hline\hline
    \end{tabularx}
    
\end{table}

\subsection{General perspective of topological band structures}\label{sec:GeneralPerspective}

We can relate our CONIs to recent progress in classifying band structures through topological obstructions of an atomic limit \cite{Bradlyn2017,Po2017}. The gluing conditions as proposed in Ref. \cite{Kruthoff2017} distinguish topological invariants of a band structure subjected to the symmetry constraints and in that sense determine all different possible configurations in momentum space that cannot be adiabatically deformed into another. This description, coinciding with $K$-theory, can be refined to determine which of these classes do not have an atomic limit. Meaning that they do not have a localized Wannier description that maintains all symmetries and are topological from this point of view. With respect to CONIs, a topological OBS implies localized OWFs cannot be constructed. In particular, one can construct symmetry indicators \cite{Po2017}. That is, the integers, such as the nine in the $p2mm$ case, define a vector space once negative entries are formally allowed, whose dimension $d_{BS}$ is given by the number of independent relations.

Heuristically, adding a band to an irrep as outlined above corresponds to moving by one unit along that ``vector". On the other hand one can determine a similar basis for all atomic insulators (AIs) in real space. Considering a specific space group, taking into account all possible position of the orbitals, or formally Wyckoff positions, results in  a similar description in terms of independent integers. These integers count how many representations are occupied which subsequently associates with a vector space, having dimension $d_{AI}$. Upon Fourier transforming, the two cases can be compared and indicators determining the absence of an atomic limit are then obtained by considering the quotient of these vector spaces. Surprisingly, $d_{AI}$ is found to match $d_{BS}$ \cite{Po2017}. Nontrivial band structures are those that cannot be written as an integer linear combination in the AI basis. On the other hand, if there exists a combination with rational coefficients that still sums up to vector in the band structure space, one concludes this is nontrivial. For example, in spinless systems with space group 106, one of the irreps always comes in an even number in AI space. One can take a rational coefficient for this vector of $\sfrac{1}{2}$, which still amounts to a vector in band structure space. This configuration has no atomic limit and is topological in this sense. 

%We do note, although this is an efficient and insightful algorithm, some invariants might map to the trivial class, in particular some instances of the tenfold way.

A related point of view is that of topological quantum chemistry \cite{Bradlyn2017}. Here as well, the different band structures arising from the gluing conditions in momentum space \cite{Kruthoff2017} are compared to atomic insulator limits in real space to isolate topological ones. In this case one departs from the language of band representations \cite{Bacry_Band}. Roughly speaking, a band representation  consists of all sets of energy bands related to localized orbitals respecting all symmetries. One can then define so-called elementary band representations (EBR) that are the smallest sets of bands derived from local atomic-like Wannier functions, serving as building blocks for this space. Contrasting the atomic case after a Fourier transform, it is then found that topological band structures, having no atomic limit, correspond to EBRs that are split in a valence and conduction band. The idea here is that then, by definition, both cannot form a band representation as one started from an EBR. It turns out, however, that with this methodology some nontrivial instances are fragile. Namely, it was uncovered that some bands might split into a fragile topological and a trivial one \cite{Po_fragile}. The fragile band arises by definition as the difference between two Wannier representable bands, i.e bands that have an atomic limit. Such topologies can therefore be trivialized upon the addition of extra bands and thus do not have stable invariants as captured by $K$-theory. The physical consequences of such represented fragile topological phases \cite{peri2020experimental, fargileclas}, including the discovery of new types that cannot be diagnosed by irreducible representation content, are still the subject of active and intense research. However, the Wilson loop spectrum \cite{Alexandrinataphase} will show winding \cite{Bouhon2018} which provides a direct diagnosis route. 

%Generically, there might be some traces of edge states in the gap although these are usually not sharp. \textcolor{orange}{Twisted boundary conditions in science in metamaterial and theory january previous year} %Especially for systems having $C_2T$, that is a combination of two fold rotations and TRS, fragile topologies have predicted. 

\section{Conclusions}

We have introduced a novel class of 2D topological matter, the optical $N$-insulators. The optical $N$-invariant counts the number of singularities in the electromagnetic linear response theory and is intimately connected to polarization transport. We have deciphered the $N$-invariant by analyzing the optical band structure of the crystal which are the eigenvectors of the atomistic susceptibility tensor. We have shown that these optical Bloch functions have obstructed Wannier representations in a nontrivial phase $N\neq 0$, thus the molecular polarizabilities lack an atomic limit. These intriguing optical $N$-phases were refined with $K$-theory to account for the underlying crystalline symmetries, providing a complete classification of the topological electromagnetic phase of matter.

\section*{Acknowledgements}
 R.-J.~S. acknowledges funding via the Marie Sk{\l}odowska-Curie programme [EC Grant No. 842901], the Winton programme as well as Trinity College at the University of Cambridge. T.~V.~M., S.~B. and Z.~J. acknowledge support from the Defense Advanced Research Projects Agency (DARPA) Nascent Light-Matter Interactions (NLM) Program and the National Science Foundation (NSF) EFRI NewLAW [Grant No. 1641101].

\appendix

\section*{Appendix}

\section{Properties of the response function}\label{app:Reponse}

Our goal is to completely classify the optical phase of any 2D material with a full spectral gap $E_\mathrm{bg}\neq 0$ and broken time-reversal symmetry. We first review some fundamental properties of optical materials that will be essential to the definition of optical invariants. The bulk material is assumed insulating and energetically stable, meaning a linear response theory is well-defined. Gapless materials $E_\mathrm{bg}=0$ like metals and semimetals are considered ``intermediate" phases which do not possess a stable topological classification in this sense. It is assumed that either a biasing magnetic field $\vec{\mathcal{B}}\neq \vec{0}$ is present to break time-reversal symmetry or there is some form of magnetic ordering.

As long as the fluctuations in the electric field are relatively weak, nonlinear effects are negligible. We also ignore magnetism $\vec{\mathcal{M}}=\vec{0}$ which is usually insignificant in optical materials. Hence, the polarization density $\vec{\mathcal{P}}$ is related to the electric field $\vec{\mathcal{E}}$ via the susceptibility tensor $\chi$ in Eq.~(\ref{eq:LinearResponseTheory}), which defines the linear response theory. We have intrinsically assumed translational symmetry in time, which means the photon energy $\omega$ is conserved. The susceptibility tensor $\chi$ is the fundamental electromagnetic quantity and completely characterizes the optical properties of the material at linear order. Note that we consider all components of the susceptibility tensor; the longitudinal, transverse and gyrotropic (Hall) responses. These components are often associated with the density-density, current-current and density-current correlations respectively. In most cases, it is the coupling between the longitudinal and transverse fluctuations that is responsible for nontrivial topologies in electromagnetism \cite{vanmechelen2020optical}.

Importantly, since electromagnetism is a real-valued field theory, the response function must satisfy the reality condition. This implies,
\begin{equation}\label{eq:reality}
\chi(\omega;\vec{r},\vec{r}')=\chi^*(-\omega;\vec{r},\vec{r}'),
\end{equation}
which is true for both dissipative and nondissipative systems. The response function is also causal and therefore satisfies the Kramers-Kronig (KK) relations,
\begin{equation}\label{eq:KK}
\oint_{\Im(\Omega)\geq 0} d\Omega\frac{\chi(\Omega;\vec{r},\vec{r}')}{\Omega-\omega}=0.
\end{equation}
$\Omega$ is a complex frequency used when we analytically continue $\chi$ into the complex plane. Equation~(\ref{eq:KK}) indicates $\chi$ is analytic in the upper half-plane and is a necessary criterion for stability of the system. The KK relations correlate Hermitian and antiHermitian components of the susceptibility tensor, as they cannot be completely independent. If the system is dissipationless, the response function is Hermitian and continuous (nonsingular) within the spectral gap $0<\hbar\omega<E_\mathrm{bg}$,
\begin{equation}
\chi(\omega;\vec{r},\vec{r}')=\chi^\dagger(\omega,\vec{r}',\vec{r}).
\end{equation}
This means the photon does not have sufficient energy to induce a transition and the medium is transparent in this region. We refer to this as the ``quantum limit", or zero temperature limit $T\to 0$, since spectral lines are point-like. The internal energy $U>0$ of the material is positive definite,
\begin{equation}\label{eq:PositiveDef}
U(\omega)=\frac{1}{2}\Re\iint d\vec{r}d\vec{r}' \vec{\mathcal{E}}^*(\omega,\vec{r})\cdot\bar{\chi}(\omega;\vec{r},\vec{r}')\cdot\vec{\mathcal{E}}(\omega,\vec{r}'),
\end{equation}
which requires that $\bar{\chi}>0$ is positive definite and defines the inner-product space,
\begin{equation}\label{eq:EnergyDensity}
\bar{\chi}(\omega;\vec{r},\vec{r}')=\frac{\partial}{\partial\omega}[\omega\chi(\omega;\vec{r},\vec{r}')].
\end{equation}
 When the excitation energy is above the band gap $E_\mathrm{bg}<\hbar\omega$, the material can absorb the photon leading to dissipation. The energy dissipation rate $Q\geq 0$ must be positive semi-definite for every $\omega$ to guarantee causality (absorption),
\begin{equation}
Q(\omega)=\omega\Im\iint d\vec{r}d\vec{r}' \vec{\mathcal{E}}^*(\omega,\vec{r})\cdot\chi(\omega;\vec{r},\vec{r}')\cdot\vec{\mathcal{E}}(\omega,\vec{r}').
\end{equation}
Note, we only consider passive media so there are no gain mechanisms present.

\section{Linear response theory (real space)}\label{app:QuantumReponse}

Here we outline the most important steps needed to extract the atomistic susceptibility tensor $\chi$ from the many-body Schr\"{o}dinger equation. This method is considered \textit{ab initio} since the only parameters that enter the theory are physical constants. More thorough derivations and computational techniques can be found in quantum transport textbooks \cite{diventra_2008,cohen_louie_2016}. We implement minimal coupling $-i\hbar\partial_\mu\to -i\hbar\partial_\mu-eA_\mu/c$ which gauges the Schr\"{o}dinger equation,
\begin{equation}
H_0\to H_0+H_I.
\end{equation}
$H_0$ is the unperturbed crystal Hamiltonian and $H_I$ is the interaction Hamiltonian that accounts for spatiotemporal fluctuations in the electromagnetic field. The crystal Hamiltonian $H_0$ is assumed to possess a complete set of Bloch states,
\begin{equation}\label{eq:BlochStates}
H_0|\psi_{n\mathbf{k}}\rangle=E_{n\mathbf{k}}|\psi_{n\mathbf{k}}\rangle,
\end{equation}
where $\mathbf{k}$ is the crystal momentum that corresponds to the center of mass motion \cite{Domenico2014}. These satisfy the Bloch condition $\hat{t}_\mathbf{R}|\psi_{n\mathbf{k}}\rangle=e^{i\mathbf{k}\cdot\mathbf{R}}|\psi_{n\mathbf{k}}\rangle$, where $\hat{t}_\mathbf{R}$ is the translation operator that shifts all particles by a lattice vector $\vec{r}_\alpha\to\vec{r}_\alpha+\mathbf{R}$. The Bloch states also have the property that $|\psi_{n\mathbf{k}}\rangle=|\psi_{n\mathbf{k}+\mathbf{G}}\rangle$, where $\mathbf{G}\cdot\mathbf{R}\in 2\pi\mathbb{Z}$ is an arbitrary reciprocal vector. Note that $n$ includes both the band and spin indices -- it is a general label for the quasiparticle excitations of the many-body system. Since our main focus is 2D materials, we assume $\mathbf{R}$ and $\mathbf{k}$ represent in-plane spatial and momentum coordinates respectively. We maintain the convention that 2D vectors are described in bold face while 3D vectors use vector arrows.

In the Weyl gauge $\phi=0$, also known as the Hamiltonian or temporal gauge, the interaction Hamiltonian $H_I$ is expressed unambiguously as,
\begin{equation}
H_I=-\frac{1}{c}\int d\vec{r}~\hat{\vec{J}}(\vec{r})\cdot\vec{\mathcal{A}}(t,\vec{r})+\sum_\alpha\frac{e^2_\alpha}{2m_\alpha c^2}\mathcal{A}^2(t,\vec{r}_\alpha).
\end{equation}
$e_\alpha$ and $m_\alpha$ are the charge and mass of the $\alpha$ particle respectively, while $\vec{\mathcal{E}}=-\partial_t\vec{\mathcal{A}}/c$ is the vector potential in the Weyl gauge. The current density operator $\hat{\vec{J}}$ is defined through the velocity operators $\vec{v}_\alpha$ as,
\begin{equation}\label{eq:CurrentDensityOperator}
 \hat{\vec{J}}(\vec{r})=\sum_\alpha\frac{e_\alpha}{2}\left[\delta(\vec{r}-\vec{r}_\alpha)\vec{v}_\alpha+\vec{v}_\alpha\delta(\vec{r}-\vec{r}_\alpha)\right].
\end{equation}
Note that the velocity operators $\vec{v}_\alpha=i[H_0,\vec{r}_\alpha]/\hbar$ may not commute due to the presence of a biasing magnetic field $\vec{\mathcal{B}}$. Fluctuations in the electric field leads to fluctuations in the internal energy. The induced current density $\vec{\mathcal{J}}_\mathrm{ind}$ quantifies the response due to these fluctuations,
\begin{equation}
\vec{\mathcal{J}}_\mathrm{ind}(t,\vec{r})=-c~\mathrm{tr}\left\{\varrho(t)\frac{\delta H_I}{\delta \vec{\mathcal{A}}(t,\vec{r})}\right\}.
\end{equation}
$\varrho(t)=\varrho_0+\delta \varrho(t)$ is the density operator and $\mathrm{tr}$ denotes the trace over the electronic Hilbert space. Here, $\varrho_0=|\psi_0\rangle\langle \psi_0|$ is the ground state density operator at zero temperature $T=0$ and $\delta\varrho$ is the density fluctuation. Note that the many-body ground state is invariant under all symmetry transformations of the crystal $\hat{t}_\mathbf{R}|\psi_0\rangle=|\psi_0\rangle$ and therefore possesses no net momentum.

It is now a fairly straightforward application of first-order perturbation theory so we will skip directly to the main result. Due translational symmetry in time it is convenient to utilize the frequency $\omega$ space. Noting the relationship between the induced current and polarization density $\vec{\mathcal{J}}_\mathrm{ind}=\partial_t\vec{\mathcal{P}}$, we obtain the atomistic susceptibility tensor,
\begin{equation}
\chi(\omega;\vec{r},\vec{r}')=-\frac{1}{\omega^2}\left[\zeta_0(\vec{r})\delta(\vec{r}-\vec{r}')\mathds{1}+\Sigma(\omega;\vec{r},\vec{r}')\right].
\end{equation}
$\mathds{1}$ is the $3\times 3$ identity in this case. $\zeta_0$ is the instantaneous diamagnetic response and $\Sigma$ is the current-current correlation function (paramagnetic response). It is important to emphasize that only the combined diamagnetic and paramagnetic response is gauge invariant. The final step is to evaluate the trace over the electronic Hilbert space. Inserting an identity $1=\sum_{n\mathbf{k}}|\psi_{n\mathbf{k}}\rangle\langle\psi_{n\mathbf{k}}|$ and using the definition of the ground state density operator $\varrho_0=|\psi_0\rangle\langle \psi_0|$ we obtain the diamagnetic function,
\begin{subequations}
\begin{equation}
\zeta_{0}(\vec{r})=\sum_\alpha\frac{e^2_\alpha}{m_\alpha}\langle\psi_0|\delta(\vec{r}-\vec{r}_\alpha)|\psi_0\rangle,
\end{equation}
and the current-current correlation function,
\begin{widetext}
\begin{equation}
\Sigma(\omega;\vec{r},\vec{r}')=\frac{1}{\hbar}\sum_{n\neq 0,\mathbf{k}}\mathrm{sgn}(\omega_{n\mathbf{k}}) \vec{J}_{n\mathbf{k}}^*(\vec{r})\otimes\vec{J}_{n\mathbf{k}}(\vec{r}')\left[\frac{1}{\omega-\omega_{n\mathbf{k}}}-i\pi\delta(\omega-\omega_{n\mathbf{k}})\right].
\end{equation}
\end{widetext}
\end{subequations}
$\otimes$ denotes the Cartesian outer product. In the last line, we have included the negative frequency spectrum in the definition of the sum, such that $n$ runs over both positive $\omega_{n}$ and negative (complex conjugate) oscillations $\omega_{-n}=-\omega_{n}$. This is necessary to preserve the reality of the response function [Eq.~(\ref{eq:reality})]. Here, $\vec{J}_{n\mathbf{k}}(\vec{r})=\langle \psi_{n\mathbf{k}}|\hat{\vec{J}}(\vec{r})|\psi_0\rangle$ is a current density matrix element and $\hbar\omega_{n\mathbf{k}}=E_{n\mathbf{k}}-E_0$ is the transition energy from the ground state $n=0$ to an excited state $n>0$. The band gap corresponds to the minimum energy $E_\mathrm{bg}=\min(\hbar\omega_{1\mathbf{k}})$ between the ground state and the first excited state, which is nonzero in an insulator. Whether the band gap is direct or indirect is not particularly important topologically, only that the spectrum is gapped for all momenta.

It is easy to check that the quantum susceptibility tensor $\chi$ satisfies all the requirements of the linear response theory. The internal energy is positive definite $U>0$, as well as the dissipation rate $Q\geq 0$. We obtain the zero temperature $T=0$ fluctuation-dissipation theorem,
\begin{equation}
Q(\omega)=\frac{\pi}{\hbar}\sum_{n\neq 0,\mathbf{k}}\left|\int d\vec{r}~ \vec{J}_{n\mathbf{k}}(\vec{r})\cdot \vec{\mathcal{E}}(\omega_{n\mathbf{k}},\vec{r})\right|^2\frac{\delta(\omega-\omega_{n\mathbf{k}})}{|\omega_{n\mathbf{k}}|}.
\end{equation}
We now have a very intuitive physical picture. When $E_\mathrm{bg}<\hbar\omega$ is above the band gap, fluctuations in the electromagnetic field $\vec{\mathcal{E}}$ leads to dissipation $Q\neq 0$ because the photon has enough energy to be absorbed and produce an electron-hole pair. However, when $0<\hbar\omega<E_\mathrm{bg}$ lies within the band gap, the system is dissipationless $Q=0$. In this regime, the photon can only polarize the material. As a consequence, the response function is Hermitian and continuous, with continuous inverse. We focus on this transparent region as it is critical for optics and the topological classification.

\section{Linear response theory (momentum space)}\label{app:MomentumSpace}

From Eq.~(\ref{eq:ResponseMomentum}), the excitation and source wave vectors are defined by $\vec{q}=(\mathbf{q}+\mathbf{G},q_z)$ and $\vec{q}'=(\mathbf{q}+\mathbf{G}',q_z')$ which conserve momentum in the plane, up to a reciprocal lattice vector $\mathbf{G}$. In the reciprocal momentum space, the diamagnetic matrix is expressed compactly as,
\begin{subequations}
\begin{equation}\label{eq:DiamagneticMomentum}
\zeta_0(\vec{q},\vec{q}')=\sum_\alpha \frac{e_\alpha^2}{V_cm_\alpha}\langle \psi_0| e^{-i(\vec{q}-\vec{q}')\cdot\vec{r}_\alpha}|\psi_0\rangle.
\end{equation}
Notice that $\zeta_0(\vec{q},\vec{q}')=\zeta_0(\vec{q}-\vec{q}')$ is independent of $\mathbf{q}$ since it only depends on the difference of wave vectors. Whereas the paramagnetic matrix is,
\begin{equation}\label{eq:ParamagneticMomentum}
\Sigma(\omega;\vec{q},\vec{q}')=\frac{1}{\hbar V_c}\sum_{n\neq 0} \frac{\mathrm{sgn}(\omega_{n\mathbf{q}})}{\omega-\omega_{n\mathbf{q}}}\vec{J}_{n}^*(\vec{q})\otimes \vec{J}_{n}(\vec{q}').
\end{equation}
\end{subequations}
$V_c$ is the 2D unit cell area of the crystal which determines the particle concentration $n_0=V_c^{-1}$. The current density matrix elements are defined as $\vec{J}_{n}(\vec{q})=\langle \psi_{n\mathbf{q}}|\hat{\vec{J}}(\vec{q})|\psi_0\rangle$, where $\hat{\vec{J}}(\vec{q})$ is the current density operator in the momentum space,
\begin{equation}\label{eq:CurrentDensityOperatorMomentum}
\hat{\vec{J}}(\vec{q})=\sum_\alpha\frac{e_\alpha}{2}  \left( e^{-i\vec{q}\cdot\vec{r}_\alpha}\vec{v}_\alpha+\vec{v}_\alpha e^{-i\vec{q}\cdot\vec{r}_\alpha}\right).
\end{equation}
Note that summation over both positive $\omega_{n\mathbf{q}}$ and negative (complex conjugate) oscillations $\omega_{-n\mathbf{q}}=-\omega_{n-\mathbf{q}}$ with all momenta reversed is implicit in Eq.~(\ref{eq:ParamagneticMomentum}). The inclusion of both positive and negative oscillators ensures the reality condition is satisfied [Eq.~(\ref{eq:Realfunction})].

The momentum space is particularly useful to implement matrix multiplication and other such operations. For example, the inverse susceptibility $\chi^{-1}$ in the Fourier basis is defined by the relation,
\begin{equation}
\sum_{\mathbf{G}''}\int \frac{dq_z''}{2\pi}\chi^{-1}(\omega;\vec{q},\vec{q}'')\chi(\omega;\vec{q}'',\vec{q}')=\mathds{1}\delta_{\mathbf{G}-\mathbf{G}'}\delta_{q_z-q_z'}.
\end{equation}
$\delta_{\mathbf{G}-\mathbf{G}'}$ is the 
Kronecker delta in the Fourier basis. Note that each instance of a matrix product involves convolution over the out-of-plane momentum $q_z$ since this is not an eigenvalue of the system. Therefore, an element of the homotopy [Eq.~(\ref{eq:Volovik})] is evaluated by,
\begin{equation}
\chi\frac{\partial\chi^{-1}}{\partial q_\alpha}=\sum_\mathbf{G''}\int \frac{dq_z''}{2\pi}\chi(\omega;\vec{q},\vec{q}'')\frac{\partial}{\partial q_\alpha}\chi^{-1}(\omega;\vec{q}'',\vec{q}'),
\end{equation}
which defines a matrix in $\vec{q}$ and $\vec{q}'$. Finally, the trace of some matrix is the convolution with itself and the summation over the diagonal,
\begin{equation}
\mathrm{tr}[\chi]=\sum_\mathbf{G} \int \frac{dq_z}{2\pi} \chi^i_i(\omega;\vec{q},\vec{q})=\sum_\lambda^{D_f} \lambda_{\omega\mathbf{q}}.
\end{equation}
Repeated indices implies summation over the polarization degrees of freedom (Cartesian tensor components).

\section{Proof of Eq.~(\ref{eq:Volovik})}\label{app:ProofEQ9}

\begin{figure}
    \centering
    \includegraphics[width=\linewidth]{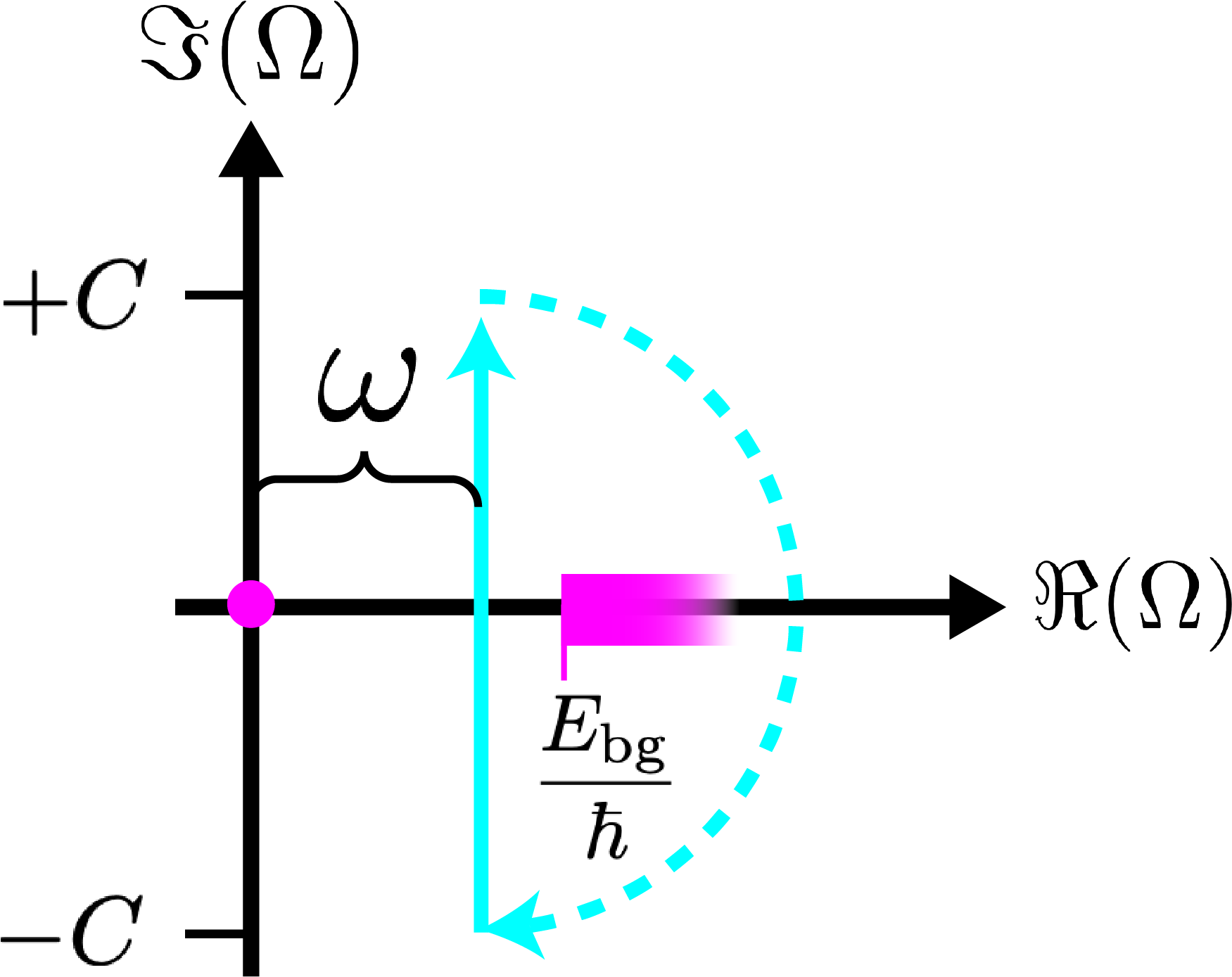}
    \caption{Contour in the complex plane to evaluate the optical $N$-invariant. $\omega$ is the photon energy that determines where the hemispherical contour is centered in the band gap. $E_\mathrm{bg}$ is the electronic band gap energy (magenta). As $C\to\infty$ the integral (cyan) is evaluated along the imaginary line and the contribution from the arc (dotted line) vanishes.}
    \label{fig:contour}
\end{figure}

An issue still remains: is the susceptibility matrix nondegenerate ($\det\chi\neq 0$) along the entire imaginary line of $\Omega$? This feature is guaranteed by the $f$-sum rule; also known as the optical, conductivity, or Thomas-Reich-Kuhn (TRK) sum rule. We follow a similar proof as Y. Zhou \& J. Liu \cite{Zhou_2020} who outline a more generalized version. In the asymptotic limit $|\Omega|\to\infty$ the susceptibility matrix approaches a purely diamagnetic response, 
\begin{equation}\label{eq:fsumRule}
\lim_{|\Omega|\to\infty}\chi(\Omega;\vec{q},\vec{q}')\to-\frac{1}{\Omega^2}\zeta_0(\vec{q},\vec{q}')\mathds{1}.
\end{equation}
The diamagnetic matrix is positive definite $\zeta_0(\vec{q},\vec{q}')>0$ and thus $\chi$ is always invertible, meaning no eigenvalues $\lambda$ are singular along this path. This is proven explicitly by taking the inner product of $\zeta_0$ with an arbitrary function in the Fourier basis $x(\vec{q})$,
\begin{equation}
\begin{split}
&\sum_{\mathbf{G}\mathbf{G}'}\iint\frac{dq_zdq_z'}{(2\pi)^2} x^*(\vec{q})\zeta_0(\vec{q},\vec{q}')x(\vec{q}')\\
&=\sum_\alpha\frac{e_\alpha^2}{V_c m_\alpha}  \langle\psi_0|\left|\sum_{\mathbf{G}}\int\frac{dq_z}{2\pi}e^{i\vec{q}\cdot\vec{r}_\alpha} x(\vec{q})\right|^2|\psi_0\rangle>0.
\end{split}
\end{equation}
Hence, the susceptibility matrix is nondegenerate $\det\chi\neq 0$ for sufficiently large contours. We can now show the integral is convergent. Applying the residue theorem on the integrand of Eq.~(\ref{eq:Volovik}) gives,
\begin{equation}
\begin{split}
-2\pi i&\sum_{\Omega_i\in \pmb{\Omega}}\mathrm{res} [\mathcal{F}(\Omega_i)]   =\oint d\Omega \mathcal{F}(\Omega)\\
&=\int_{\omega-iC}^{\omega+iC}d\Omega\mathcal{F}(\Omega)+\int_\mathrm{arc}d\Omega\mathcal{F}(\Omega),
\end{split}
\end{equation}
where $\mathcal{F}=\epsilon^{\alpha\beta\gamma}\mathrm{tr}[A_\alpha A_\beta A_\gamma]$ and $A_\alpha=\chi\partial_\alpha\chi^{-1}$ is an element of the homotopy. The arc radius $C$ is large enough to encircle all poles $\Omega_i\in \pmb{\Omega}$ in the right complex plane, where $\pmb{\Omega}$ is the set of poles bounded by the contour. An illustration of the complex contour is shown in Fig.~\ref{fig:contour}.

For large enough contours $C\to\infty$, the temporal component $A_\Omega$ is proportional to the identity multiplied by a simple pole $A_\Omega\propto 1/\Omega$ due to the $f$-sum rule [Eq.~(\ref{eq:fsumRule})]. As $C\to \infty$, the arc integral is then proportional to,
\begin{equation}
\int_\mathrm{arc}d\Omega\mathcal{F}(\Omega)\propto \int_\mathrm{arc}\frac{d\Omega}{\Omega}\epsilon^{ij}\mathrm{tr}[A_iA_j]\to 0,
\end{equation}
where $A_{i}$ are the spatial components. The arc integral vanishes due to antisymmetry in $\epsilon_{ij}=-\epsilon_{ji}$ and the cyclic property of the trace. The imaginary line integral is convergent,
\begin{equation}\label{eq:ShiftProperty}
\begin{split}
&-2\pi i\sum_{\Omega_i\in \pmb{\Omega}}\mathrm{res} [\mathcal{F}(\Omega_i)]=\int_{\omega-i\infty}^{\omega+i\infty}d\Omega\mathcal{F}(\Omega)\\
&=\int_{-i\infty}^{i\infty} d\Omega\mathcal{F}(\Omega+\omega)=i\int_{-\infty}^{\infty} d\tau\mathcal{F}(\omega+i\tau),
\end{split}
\end{equation}
which defines the homotopy of a circle $S^1$ when including the point at $|\Omega|=\infty$. In the last line, we have shifted the integral to the imaginary axis $\tau\in (-\infty,\infty)$. This property will aid us for an important proof in App.~\ref{app:ProofEq10}.

The final step to prove $N\in\mathbb{Z}$ is quantized requires Bloch periodicity in $\mathbf{q}$. We consider an arbitrary deformation in the response function $\chi\to \chi+\delta\chi$. The corresponding variation in the invariant $N\to N+\delta N$ is,
\begin{equation}
\delta N=-\frac{\epsilon^{\alpha\beta\gamma}}{8\pi^2}\iint  d\Omega d\mathbf{q}~\partial_\alpha\mathrm{tr}\left[\delta\chi\partial_\beta\chi^{-1}\chi\partial_\gamma\chi^{-1}\right].
\end{equation}
All variations amount to a total divergence in the integrand. We have already proven that the temporal boundaries vanish as a consequence of the $f$-sum rule. The spatial components are zero as well due to the periodic boundary condition $\mathbf{q}\to\mathbf{q}+\mathbf{G}$. Hence, all variations vanish identically $\delta N=0$ and $N\in\mathbb{Z}$ is topologically quantized.

\section{Homotopy equivalence between susceptibility and conductivity matrices}\label{app:CondHomotopy}

It should be emphasized that the topology is not predicated on our choice of the susceptibility tensor $\chi$ over the conductivity tensor $\sigma$. The optics community prefers $\chi$ while $\sigma$ is used more often in condensed matter. Here we prove that $N$ is unchanged under the substitution of the conductivity tensor $\chi=i\sigma/\Omega$, which may be used in place of $\chi$ to calculate the optical $N$-invariant. Under this substitution we get $N\to N+\delta N$,
\begin{equation}
\delta N=\frac{-1}{8\pi^2}\int_{\omega-i\infty}^{\omega+i\infty} \frac{d\Omega}{\Omega}\int d\mathbf{q}~\epsilon^{ij}\partial_i\mathrm{tr}\left[\sigma\partial_j\sigma^{-1}\right]=0,
\end{equation}
where $N$ is equivalently defined through $\sigma$,
\begin{equation}
N=\frac{\epsilon^{\alpha\beta\gamma}}{24\pi^2} \iint d\Omega d\mathbf{q}~\mathrm{tr}\left[\sigma\frac{\partial\sigma^{-1}}{\partial q_\alpha}\sigma\frac{\partial\sigma^{-1}}{\partial q_\beta}\sigma\frac{\partial\sigma^{-1}}{\partial q_\gamma}\right].
\end{equation}
Since $\delta N$ is a total derivative, the surface terms vanish due to the periodic boundary condition in $\mathbf{q}$. Thus, the susceptibility and conductivity tensors are homotopy equivalent.

\section{Proof of Eq.~(\ref{eq:BerryCurvature})}\label{app:ProofEq10}

Our goal is to simplify the expression in Eq.~(\ref{eq:Volovik}) to avoid temporal integration and matrix inversion. We follow a similar procedure as Z. Wang \& S.-C. Zhang \cite{Wang2012,ZhongWang2012} which walks through the main steps of the proof. For consistency, it is useful to define the Hermitian part of the conductivity tensor $\varsigma=-i\sigma=\Omega\chi$ since it has purely real eigenvalues along $\Im(\Omega)=0$ and strictly first-order poles. Indeed, using a partial fraction decomposition, we express the $\varsigma$ matrix in standard Green's function form \cite{Gurarie2011,Zhou_2020},
\begin{equation}\label{eq:PartialFraction}
\varsigma(\Omega;\vec{q},\vec{q}')=-\frac{1}{\Omega}\zeta(\vec{q},\vec{q}')-\frac{1}{\hbar V_c}\sum_{n\neq 0} \frac{\vec{J}_{n}^*(\vec{q})\otimes \vec{J}_{n}(\vec{q}')}{|\omega_{n\mathbf{q}}|(\Omega-\omega_{n\mathbf{q}})}.
\end{equation}
$\zeta$ is the gauge invariant diamagnetic response that only responds to transverse fields $\vec{q}\cdot\vec{\mathcal{E}}=0$, 
\begin{equation}\label{eq:StaticCurrentCurrent}
\zeta(\vec{q},\vec{q}')=\zeta_0(\vec{q},\vec{q}')\mathds{1}+\Sigma(0;\vec{q},\vec{q}').
\end{equation}
Note that $\zeta>0$ is positive definite, which ensures the energy density [Eq.~(\ref{eq:EnergyDensity})] is positive $\bar{\chi}=\partial_\omega\varsigma>0$ for all $\omega$.

It was proven in App.~\ref{app:CondHomotopy} that the homotopy of $\varsigma$ and $\chi$ are equivalent so we may use $\varsigma$ in place of $\chi$ to calculate the $N$-invariant. Now consider the continuous deformation in $\varsigma$ characterized by the smooth homotopy $0\leq\epsilon\leq 1$,
\begin{equation}
\varsigma_\epsilon(\omega,i\tau;\vec{q},\vec{q}')=(1-\epsilon)\varsigma(\omega+i\tau;\vec{q},\vec{q}')+\epsilon\tilde{\varsigma}(\omega,i\tau;\vec{q},\vec{q}').
\end{equation}
We exploit the shift property in Eq.~(\ref{eq:ShiftProperty}) to redefine the integral along the imaginary line $\tau\in (-\infty,\infty)$. The unitless paramater $\epsilon$ interpolates between the true response function $\varsigma$ at $\epsilon=0$ and the modified response function $\tilde{\varsigma}$ at $\epsilon=1$,
\begin{equation}
\tilde{\varsigma}(\omega,i\tau;\vec{q},\vec{q}')=\left[-i\tau+\varsigma^{-1}(\omega;\vec{q},\vec{q}')\right]^{-1}.
\end{equation}
$\tilde{\varsigma}$ has nonzero eigenvalue along $\tau\in (-\infty,\infty)$ and any sufficiently large contour in the complex plane $|\Omega|\in C$. It is also clear that $\varsigma_\epsilon(\omega,0;\vec{q},\vec{q}')=\varsigma(\omega;\vec{q},\vec{q}')$, so the deformation is purely in the complex plane. From App.~\ref{app:ProofEQ9}, we know that any nondegenerate deformation in $\varsigma$ is permitted. We only have to prove that the eigenvalues of $\varsigma_\epsilon$ never cross zero for every $0\leq\epsilon\leq 1$. This is confirmed by contracting $\varsigma_\epsilon$ with an arbitrary vector $\vec{x}(\vec{q})$ and considering the imaginary part,
\begin{equation}
\begin{split}
 &\sum_{\mathbf{G}\mathbf{G}'}\iint\frac{dq_zdq_z'}{(2\pi)^2}\Im[\vec{x}^*(\vec{q})\cdot\varsigma_\epsilon(\omega,i\tau,\vec{q},\vec{q}')\cdot\vec{x}(\vec{q}')]\\
&=(1-\epsilon)\tau\Bigg\{\frac{\sum_{\mathbf{G}\mathbf{G}'}\iint\frac{dq_zdq_z'}{(2\pi)^2}\vec{x}^*(\vec{q})\cdot\zeta(\vec{q},\vec{q}')\cdot\vec{x}(\vec{q}')}{\tau^2+\omega^2}\\
&+\frac{\left|\sum_{n\neq 0,\mathbf{G}}\int\frac{dq_z}{2\pi}\vec{J}_n(\vec{q})\cdot\vec{x}(\vec{q})\right|^2}{|\omega_{n\mathbf{q}}|\left[\tau^2+(\omega-\omega_{n\mathbf{q}})^2\right]}\Bigg\}\\
&+\frac{\epsilon\tau }{\tau^2+\left|\sum_{\mathbf{G}\mathbf{G}'}\iint\frac{dq_zdq_z'}{(2\pi)^2}\vec{x}^*(\vec{q})\cdot\varsigma^{-1}(\omega,\vec{q},\vec{q}')\cdot\vec{x}^*(\vec{q}')\right|^2}.
\end{split}
\end{equation}
Since $\zeta>0$, the imaginary part is nowhere zero in the complex plane. It is always positive or negative in the upper $\tau>0$ or lower $\tau<0$ regions, respectively. Hence, the matrix $\varsigma_\epsilon$ is never degenerate for every $0\leq\epsilon\leq 1$.

As a consequence, the optical $N$-invariant can be computed from the modified Green's function $\varsigma_1=\tilde{\varsigma}$, which has a much simpler frequency dependence. Utilizing the spectral decomposition of the optical band structure [Eq.~(\ref{eq:SpectralDecomp})] we arrive at,
\begin{equation}
\begin{split}
N&=\frac{i}{4\pi^2}\sum_\lambda^{D_f} \int d\mathbf{q} \int_{-\infty}^{\infty} d\tau\frac{\partial}{\partial \tau}\log \left[-i\tau+(\omega\lambda_{\omega\mathbf{q}})^{-1}\right]\mathfrak{F}_\lambda\\
&=\frac{1}{4\pi}\int d\mathbf{q}\sum_{\lambda}^{D_f}\mathrm{sgn}(\lambda_{\omega\mathbf{q}}) \mathfrak{F}_\lambda=\frac{1}{2\pi}\int d\mathbf{q}~\mathfrak{F},
\end{split}
\end{equation}
where $\mathfrak{F}_\lambda$ is the Berry curvature of a single OBF $\vec{e}_\lambda$ and $\mathfrak{F}=\sum_{\lambda>0}\mathfrak{F}_\lambda$ is the sum over all capacitive modes $\lambda>0$. Note that one can also calculate the curvature from the inductive modes $\mathfrak{F}=-\sum_{\lambda<0}\mathfrak{F}_\lambda$ as the total summation vanishes $\sum_{\lambda}^{D_f}\mathfrak{F}_\lambda=0$.

\bibliography{top_optics.bib}

\end{document}